\documentclass[aps]{article}
\usepackage{geometry}                
\usepackage{graphicx}

\usepackage{amssymb}
\usepackage{epstopdf}
\usepackage{amsmath}

\usepackage{amsfonts}
\usepackage{bm}
\usepackage{hyperref}

\DeclareMathAlphabet{\mathpzc}{OT1}{pzc}{m}{it}

\title{Globally Causal Solutions for Gravitational Collapse}
\author{Clifford E Chafin\\\ \small{Department of Physics, North Carolina State University, Raleigh, NC 27695} \thanks{cechafin@ncsu.edu}}

\begin{document}
\maketitle

\begin{abstract}
Through an illuminating thought experiment we demonstrate that the nonsingular ``continued collapse'' picture of a black hole is the only consistent and physical one.  We provide a class exact solutions on the boundary of the space of physical configurations.  This will show that all the other known exact solutions are unphysical near the surface of the event horizon or in the interior.  
This will have important consequences for the ``no-hair'' conjecture and the kinds of persistent fields that can emerge from a black hole as well as the evolution during collisions and near grazing events.  The interior of these holes are characterized by a limiting degenerate metric and these regions tend to well defined volumes and radii in contrast with what is inferred from singular solutions.  Surprisingly, these depend on past history and not simply the mass or external fields of the body.  It is shown that there is often a well defined ``hidden'' flat background that can be used to equivalently reformulate GR in terms of a classical nonlinear gravity field and gives local conservation laws.  This has implications for unification efforts and numerical approaches to handle the degenerate metric regions reminiscent of the Rankine-Hugoniot conditions.  Possible consistency problems with current numerical approaches to black hole dynamics are discussed.  
\end{abstract}

The history of geometry and dynamics naturally meet in general relativity (GR).  In fact, the development of the theory of invariants seem to have inspired early work on relativity and notions of gauge freedom.  In one direction, one has the geometrodynamic approach \cite{MTW} where invariance is at the forefront and a manifold structure is presumed.  Since this divided gravity at a fundamental level from the other fields of nature, unification efforts have alternately proceeded by treating GR as a perturbation on a flat background \cite{Weinberg}.  This has the limitation that strong curvature systems are not clearly attainable by such a method.  

Our goal here is to first show that consistency requires that only a limited set of ``causally connected'' evolution paths be chosen.  This will specifically rule out the formation of any singularities at any time and will make the use of Lagrangian observer proper time arguments sometimes invalid.  This will have important implications for the ``no-hair'' conjecture and show that the kinds of fields we can expect from black holes can be significantly more elaborate.  The picture of a black hole as a ``frozen star'' is an old one but here shown to be essential and not equivalent to the usual singular ones in crucial ways.  We present a class of exact nonrotating infalling solutions that retain the full history of their creation and demonstrate that, unlike the results derived from singular solutions, that the radii and volume are well defined although, surprisingly history dependent, so not a simple function of the black hole mass.  

This frozen star picture is especially problematic when it comes to numerical solutions because the metric degenerates.  The ADM decomposition of the Einstein equations are not amenable to such a picture so we need some other tools to approach the problem.  In the case of shock waves, conservations laws allow the derivation of the Rankine-Hugoniot conditions to handle the discontinuity at the shocks.  We will show that one can create an underlying structure to rewrite GR as a classical field theory on this ``hidden'' flat background that allows true local conservation laws to exist and resolve problems with variational methods.  This may provide a method to treat the degenerate boundary\footnote{Even though we will see the horizon never truly forms, there is a convergence to a degenerate region and the limiting boundary of it we may casually refer to as the ``event horizon.''} and its interior for numerical analysis of such black holes.  The existence of such a structure raises questions about the relativity principle itself and allows a generalization of it which suggests an approach to unification that treats gravity on more of a par with the other fields.

\section{Discussion}
The meaning of Einstein's equations has been a subject of discussion since they were first written down.  Unlike most other classical field equations, the timelike and spacelike aspects of the universe are not trivially and persistently decomposable but are altered by the presence of sources and the gravitational field itself.  Since we will be proposing a new way of assigning meaning to the equations we give a short historical perspective on how our modern view of dynamics and coordinates arose.  Galileo in many ways is responsible for the notion of vectors and coordinates in physics.  His ideas anticipate the later work of Descartes in coordinatizing geometry.  Additionally, he
first introduced time as an independent variable  and so essentially founded dynamics.  The picture of space and time as a movie of ``snapshots'' of a 3D world indexed by time is implicit in Newtonian mechanics and was only updated by Einstein's realization that the velocity of light must be independent of observer.  This required a mixing of space and time variables in different ``reference frames'' of inertial observers and forms the basis of special relativity.  Later, it was realized that one can consider a more general foliation of space and time into general space-like slices.  None of this requires curvature be present but it provides a powerful starting point for the set of curvature based models of gravity that includes general relativity (GR).  

In special relativity, we typically seek equations of motion and invariants that are ``manifestly covariant.''  Specifically, the equations should be written in a tensor form where the invariance under boosts is trivial.  In the case of curvilinear coordinates, the framework of differential geometry allows one to simply elevate the form of these by transforming coordinate derivatives to covariant derivatives.  The explicit results are infamously complicated but, once a metric is provided, the mechanics are clear.  Fundamental equations in classical (nonquantum) physics are typically hyperbolic.  These incorporate the generally second order kinematics of nature and the strict causal behavior implicit in the realization that lightspeed is the maximum velocity of propagation.  

Once we have the Einstein equations, we require a manifold for their evolution (and, usually, a specified differential structure if the dimensionality is greater than three).  It is usually implicitly assumed that spacetime is a 3D manifold with a 1D time that is noncompact.  From here we either look for exact solutions or use a 3+1 Hamiltonian decomposition of the equations in terms of data on a manifold of 3D spacelike slices.  As long as every point in the manifold at one time can be causally reached in some finite time from every other point there is no ambiguity.  Problems arise for singularities.  These arose historically in the study of ``black holes.''  To get a sense of why these cause difficulty, one should remember why the general coordinates are so different in the case of curvature.  We can write down some exact solutions with singularities like the Schwarzchild, Eddington-Finkelstein, Kruskal and others.  These are all interpreted as representing the same physical reality in different coordinates.  We might choose to use labels like $(x,y,z,t)$ at one point where the $t$-coordinate is time-like and the others space-like, however, as we try to extend these coordinates this property may not be able to be maintained.  This problem never has to arise in flat spacetime.  

In the early days of the subject, it was realized that one could have solutions where not even light could escape.  The Schwarzchild solution gives singularities at the boundary where this begins and at the origin.  Eventually, it was recognized that the singularity at this boundary is not physical but the one at the origin is and corresponds to an infinitely dense region of finite mass.  This suggests the modern view.  Beyond a certain density the Lagrangian picture of inertial observers gives a set of ``trapped light surfaces'' which indicate that the formation of a black hole is irreversible.  This seems to suggest that the final state will be a point-like mass surrounded by an ``event horizon'' defining the limit of possible outwards propagation from it.  The symmetries of such a solution lead to the famous ``no-hair'' conjecture which dictates that the only surviving properties of a black hole are mass, angular momentum and charge.  The magnetic field, in particular, must vanish.\footnote{  Numerical approaches incorporating this perspective always involve some creativity in dealing with this singularity.  One could naturally ask how information of changing external fields can induce forces on this singularity given that changing fields can reach it but it cannot send signals out to the event horizon to tell it how much to move.}  

This view was not always held.  Many thought that small symmetry violations in the initial data would cause a true singularity never to form.  The Hawking and Penrose theorems \cite{Hawking} are considered to show that this is not the case and, after a certain point, violations of spherical symmetry are unimportant and the ``continued collapse'' to a singular state will occur in finite proper time for the local observers.
Early thoughts on black holes included the notion that infalling matter never makes it to the center because of redshifting due to the increasing effects of gravity.  This view is not the one that has gained traction, probably because of the heavy reliance on exact solutions and the dominance of arguments based on freely moving observers along geodesics.  

The purpose of this article is to argue that the Lagrangian point of view is generally inconsistent.  To justify this consider the following thought experiment.   Consider a ball of dust that is falling in on itself.  Since it generates no pressure, collapse is inevitable.  It is well known, for observers far away, the particles are redshifted to near black and  approach the speed of light.  Their local clocks seem to be arrested and they take an infinite amount of time to form an event horizon.  In this picture the singularity and horizon never form.  Every particle and point of space remains causally (forwards and backwards) connected to every other point in space.  What is not sufficiently well appreciated is that the formation of a black hole is so stalled at every level and that its ``frozen'' state can be one or more shells or a continuum of them with a nonsingular interior.  This has implications for enduring (and retrievable!) internal structure and its external field.  

Consider the case of a particular set of initial data or conspiring alien civilization that seeks to extract mass from this infalling cloud by moving much larger and denser bodies near it so that some of the particles leave the collection and follow and, possibly, join the larger mass.  The aliens have infinite time to pursue this project and so can, conceivably, remove an arbitrarily large fraction of mass from it.  If we embrace the point of view that the Lagrangian pictures dictates the singularity must form we then must face this contradiction.  We are implicitly assuming that no such interference ever happens, ergo, it assumes something about the entire (possibly infinite) history of the external universe.  In contrast, a global Eulerian approach assumes a global spacelike slice exists for the initial data and for all times as measured far from the cluster.  We see that it gives forwards and backwards causal connection between any pair of points for all times (however this is locally defined).\footnote{It is a bit ambiguous how to specify two such points on an evolving manifold.  The metric and coordinates can change so how we define the \textit{same} spatial point at different times in an unambiguous or physical manner is unclear and maybe impossible.  However, since we require it of all pairs of points and all times, any specification should work.  This is clear for a compact manifold.  For the noncompact case, we consider the space to be asymptotically flat with the point assignments far away to be constant.  This reduces the problem to the compact case.  Further discussion of such points is relegated to the appendix.}  Local evolution in the mass cluster drags to a halt as the local clocks are slowed to zero relative to the clocks far away.  

It seems that the fundamental notion of relativity, that all observers are equivalent, is violated here.  The falling observers do not have the same status as the observers far from the forming body.  It is the external observers in the universe that can nix the formation of the singularity, or alter its size, in their temporally unbounded futures.  On the other hand, one can state this as ``an event that is observed by one observer is observed by all observers.''  This might be a better generalization of the principles of special relativity to the general case than the usual notion that all Lagrangian histories are valid for all values of their proper time.  From a numerical perspective, this implies that, if the initial data exists in one global spacelike frame coverable in one chart, then it is so for all times.  Lapse and shift choices must be selected to enforce this, although the metric can become arbitrarily close to degenerate and rotation can cause the cones to become so tilted that the $\hat{t}$ directions may become spacelike as well (as in the erogoregion of the Kerr solution).  However, the forward timelike cone remains entirely in the positive $t$-halfspace so that the advancing slices indexed by $t$ does give a meaningful sense of the future.  
 
As a consequence, all of the singular solutions of general relativity are unphysical for a universe that starts with no such singularities.  We should only consider in the set of allowable physical data solutions that do not include them (since they are not relevant and too presumptuous about the actions of those of us on the outside).  We will present a closed form limiting solution arbitrarily close to physical black hole solutions that reveal how the history of infalling matter and charge is preserved and allow strong magnetic fields and electric multipoles to persist for all time.  The previous argument suggests that black hole interactions can even liberate matter in reverse order of its accretion that would usually be considered to be lost forever.  

The absence of such singular solutions resolves (or renders irrelevant) many problems such as the ``cloaking'' problem of singularities, the destruction of entropy and, possibly, the problem of how to build a consistent field theory for gravity.  It seems the results \cite{Hawking} on the inevitability of black hole formation are rendered meaningless.  
It will turn out that we can now consistently formulate a theory of gravity on a flat background where the ``geometrodynamic'' features are replaced by an alternate nonmetric field that controls the evolution of ``clocks'' and displacement measures of ``rods.''  More precisely, the evolution equations of the fields $A^{\mu}$ and others will be modified in a nonlinear fashion in terms of a nonmetric indexed field $h^{\mu \nu}$.  This will require some care so that the usual tensor index rules don't lead to confusion.  An end result is that we will have a set of conservation laws with true local meaning.  This does not exist in usual general relativity where the best one obtains is the York-Brown quasilocal action \cite{York}.  The methods to compute this in terms of the ``observables'' is not necessarily trivial but its existence is still of value for checks on numerical results and to place bounds on the kinds of evolution that are possible; and enduring problem in traditional GR.  

Part of the usual reasoning in deriving GR is that the equations of motions should be explicitly ``coordinate independent.''  This assures that the physics never depend on something so arbitrary.  We can, however, generalize further.  One \textit{can} allow coordinate dependence if the fields conspire in such a way that the local reality of an object we derive from the physical fields gives an equivalent class of evolved systems.  Essentially, this is a kind of gauge invariance.  We simply state it this way to express that the reality of the physics we observe may be more entangled among the fields than we are used to reading off in terms of their conserved and other gauge invariant quantities.  We might distinguish these two cases as ``coordinate relativity'' and ``field relativity'' where the second case allows fields to vary with coordinates but such that they give a net ``reality'' that is not dependent on them.  An analog would be the case of the A-B effect where the observable reality $(E, B, \rho, v)$ of fields and particles are such that the velocity of the particle is not entirely encoded in the descriptor $\psi$ that we associate with it but is a function of it and the vector potential $A$ through $\rho v=j=-\Re\psi^{\dagger}\nabla\psi+eA\psi^{\dagger}\psi$.   
We suggest that to give a uniform treatment of gravitational field with the other fields of nature, the ``field relativity'' approach is essential.  
This does not involve a change in Einstein's equations but it does involve interpreting and decomposing the fields differently and in evolving them in a less general fashion than is currently done in numerical simulations.

\section{The Black Hole Time-Frozen Solution}
We will see here that the natural limit of a nonrotating spherically symmetric dust cloud with smooth and slowly varying initial velocity field is particularly simple.  It gives a gravitating system that is described by a decomposition into two regions: an interior solution with a degenerate metric of infalling matter where the rate of time advance has vanished and an exterior ``isotropic'' vacuum solution familiar from many classical black hole results.  In order to avoid true degeneracy in our calculations, we investigate the case where the time component of the metric satisfies $g_{tt}=\epsilon(x)$ inside the event horizon.  A more realistic result may have it have a more specific spatial dependence and evolution will slowly alter this further but knowing that it is small is sufficient for our purposes.  In our example below, $g_{tt}$ indicates the local evolution rate of the system so, as it vanishes, we consider it to be ``time-frozen.''  The solution tends to become static everywhere but, in this region, a clock (or other local perturbation) appears fixed in time.  

In such a degenerate limit, the components of tensors with raised and lowered indices, e.g.\ $T^{\mu\nu}$ and $T_{\mu\nu}$, become very different and separately keeping track of these limiting cases is essential.  Accordingly, we will first discuss the details of the stress energy tensor from the Eulerian point of view in the time-frozen limit.  This will then give a general interior solution with rich possibilities for history dependence even in this simple rotationally invariant static limit.

\subsection{The Stress-Energy Tensor}

The microscopic stress-energy tensor in flat space is given in coordinate form by $\rho u^{\mu}u^{\nu}$ where $u^{i}=dx^{i}/dt$ where $x^{i}$ are linear spacelike coordinates and $t$ is a linear timelike one.  The value $u^{t}$ is fixed by the condition $g_{\mu\nu}u^{\mu}u^{\nu}=1$ and $\rho$ is the mass density of a single particle or collectively moving parcel of such particles.  
We will be only considering diagonal metric tensors in what follows.  This is equivalent to stating that the light cones do not ``tilt.''  In the Eulerian picture we need to have some way to discuss the relation of the coordinate velocity to the boost parameters that will bridge the components of the stress tensor with the underlying particle and mass density.  

The conceptually safest treatment of point particles in a field theory is via wavepackets from hyperbolic pdes.  These can be written $g^{\mu\nu}\partial_{\mu}\partial_{\nu}\Phi+\ldots$ so that the local causal structure is determined by $g$.  
We are interested in a picture where the time rate of evolution of the slices is the same for all positions in space.\footnote{To even use the term ``positions in space'' in GR is to adopt a profoundly Eulerian and nongeometric perspective.  The meaningfulness of such ``points'' will be discussed in the appendix.}  This is so that we can view the evolution as a trivial foliation of space and time just as in the flat case.  From the metric we can extract the local coordinate velocity of light by using $dx^{i}$ and $dt=dx^{0}$ for a null surface.  The radial \textit{coordinate} velocity of light is given by $c^{2}(x)=g_{tt}/g_{rr}$ so we can assign the boost factor $\gamma(v,c)=(1-v^{2}(x)/c^{2}(x))^{-\frac{1}{2}}$ for a massive particle moving with (radial) coordinate velocity $v$.  Since we plan to evolve the space as global slices advancing with uniform temporal rate $v^{t}=1$ this gives the correct boost factor for any particle moving in this local light cone.  

We are going to be especially interested in the ``time-frozen'' limit where $g_{tt}\rightarrow0$ because this corresponds to the case of a collapsing object as seen by observers at infinity.  Such a metric tends to degeneracy and our solutions will be for small but nonzero values of $g_{tt}=\epsilon(x)$.  These will be ``nearly static'' solutions.  These nearly, but still not degenerate, cases are important because we want to know what the stress tensor is in raised and lowed index form and the limiting details contain essential information.  The Einstein equations are most naturally written in lowered index form $G_{\mu\nu}=8\pi T_{\mu\nu}$ but the raised $T^{tt}$ component is what is integrated to compute the mass density.  

The stress tensor can now be generally defined as $\rho u^{\mu}u^{\nu}$ where $u^{t}=\gamma(v,c)$ and $u^{i}=\gamma(v,c)v^{i}$.  As we approach the time-frozen limit we see that $c\rightarrow0$ and yet the boost factors can still be very large even though the coordinate velocities $v^{i}$ also vanish.  To compute the actual mass in a region we evaluate $\int d^{3}x T^{tt}\sqrt{|g_{ij}|}$  The determinant of the spatial part of the metric (indicated here by the presence of only roman indicies) remains finite but the value $T^{tt}$ can be large either because the mass density $\rho$ is large or the boost factor $\gamma$ is so.  Since the coordinate velocity tends to vanish independent of history we see that the details of the history of formation is crucial in determining if the boost factor is large, hence how much mass or how many particles went into creating a given sized source of the gravitational field.   

We are going to be particularly interested in diagonal nearly degenerate metrics of the form
\begin{equation}
g_{\mu\nu}=
\begin{pmatrix}
\epsilon & 0 & 0 & 0 \\ 0 & g_{rr} & 0 & 0\\0 & 0 & g_{\theta\theta} & 0\\0 & 0 & 0 & g_{\phi\phi}
\end{pmatrix}
\end{equation}
Stress-energy tensors corresponding to radial smooth dust flows take the form
\begin{equation}
T^{\mu\nu}=
\begin{pmatrix}
\rho\gamma^{2} & \rho\gamma^{2}v^{r} & 0 & 0 \\ \rho\gamma^{2}v^{r} & \rho\gamma^{2}v^{r}v^{r} & 0 & 0\\0 & 0 & 0 & 0\\0 & 0 & 0 & 0
\end{pmatrix}
\end{equation}

We can consider two distinct cases.  Firstly, we assume that we lower mass ``adiabatically'' to form the horizon on filaments.  (Later, we consider the case of free infall from infinity.)  
In this case the stress tensors come from the $v^{r}=0$ case (so that $T^{\mu\nu}=\text{diag}[\rho,0,0,0]$) so that $\gamma\sim1$.  The radial speed of light in this region is $c=\sqrt{g_{tt}/g_{rr}}\sim\epsilon^{\frac{1}{2}}$ so all radial velocities must be much smaller than this.  The \textit{lowered} index form of the stress tensor is what appears in the Einstein equation:
 \begin{equation}
T_{\mu\nu}=
\begin{pmatrix}
\epsilon^{2}\rho\gamma^{2} & \epsilon\rho\gamma^{2}g_{rr}v^{r} & 0 & 0 \\ \epsilon\rho\gamma^{2}g_{rr}v^{r} & \rho\gamma^{2}g_{rr}^{2}v^{r}v^{r} & 0 & 0\\0 & 0 & 0 & 0\\0 & 0 & 0 & 0
\end{pmatrix}
\end{equation}
Assuming $\mathcal{O}(g_{rr})\sim1$ and $\gamma\sim1$ then the stress tensor obeys
 \begin{equation}
T_{\mu\nu}\le
\begin{pmatrix}
A\epsilon^{2}& B\epsilon^{\frac{3}{2}} & 0 & 0 \\ B\epsilon^{\frac{3}{2}}& C\epsilon & 0 & 0\\0 & 0 & 0 & 0\\0 & 0 & 0 & 0
\end{pmatrix}\approx 0
\end{equation}
This shows that we can use the vacuum field equations in seeking nearly static solution for such a mass distribution.  
We see that $T^{tt}\approx 0$.  Assuming particle number of the infalling particles are preserved how can this be?  The filaments carry of the gravitational potential energy\footnote{Potential energy is not a well defined concept in GR but we will make it more precise in our later discussion.} of the particles out to infinity.  By the time they reach the horizon this accounts for \textit{all} the mass energy they possess.  As such, the critical radius vanishes and yet contains an arbitrarily large number of particles with no net mass!  It is good to remember from radiation reaction discussions \cite{Rhorlich} that the ``mass'' of a particle incorporates its delocalized electromagnetic and gravitational fields and is computed from a case where it is isolated in flat space so that nonlocal and nonlinear alterations of the field quite reasonably alter its mass.  

Now we consider the more physically reasonable case of the black hole formed by particles (with some possible initial kinetic energy) falling in from infinity.  The large $\gamma$ factors and small value of $\epsilon$ now ensure that $T_{rr}$ dominate
 \begin{equation}
T_{\mu\nu}\approx
\begin{pmatrix}
0 & 0 & 0 & 0 \\ 0 & T_{rr} & 0 & 0\\0 & 0 & 0 & 0\\0 & 0 & 0 & 0
\end{pmatrix}
\end{equation}
from which we can construct an interior solution.  The single nonzero component incorporates all the information we have in the time frozen limit of the mass and kinetic energy distribution of the sources that manifest in the external gravitational field.

\subsection{Interior and Exterior Solutions}

The isotropic coordinate solution has the advantages that it has no spurious singularities at the horizon and it is diagonal and so the light cones do not tilt relative to the $t$-coordinate.  Since we want a rotationally symmetric static solution and, for physical reasons, we are only interested in solutions that have global causal connection for all time, a solution with all ``vertical cones'' is necessary.  If the cones tilted inwards, the source material could continue inwards and form a horizon.  The cones in the support of the matter sources must become time frozen in the limit else the matter would continue inwards or outwards.  For the external field the isotropic coordinates give such a vacuum solution.  

To find a class of allowable interior solutions we consider the set of time-frozen metrics
\begin{equation}
g_{\mu\nu}^{(\text{inner})}=
\begin{pmatrix}
\epsilon & 0 & 0 & 0 \\ 0 & -f(r) & 0 & 0\\0 & 0 & -r^{2}h(r) & 0\\0 & 0 & 0 & -r^{2}\sin^{2}(\theta)h(r)
\end{pmatrix}
\end{equation}
where we have built in an angular isotropy.  The general solution must be of the form
\begin{equation}
g_{\mu\nu}^{(\text{inner})}=
\begin{pmatrix}
\epsilon & 0 & 0 & 0 \\ 0 & -F(r) & 0 & 0\\0 & 0 & -C & 0\\0 & 0 & 0 & -C\sin^{2}(\theta)
\end{pmatrix}
\end{equation}
Einstein's equations have the form
\begin{align}
G_{\mu\nu}=R_{\mu\nu}-\frac{1}{2}g_{\mu\nu}R=8\pi T_{\mu\nu}
\end{align}
Solving for the Einstein tensor
\begin{equation}
G_{\mu\nu}^{(\text{inner})}=
\begin{pmatrix}
-\frac{\epsilon}{C} & 0 & 0 & 0 \\ 0 & \frac{F(r)}{C} & 0 & 0\\0 & 0 & 0 & 0\\0 & 0 & 0 & 0
\end{pmatrix}
\end{equation}
This seems to give $F(r)=8\pi C\rho\gamma^{2}g_{rr}^{2}v^{r}v^{r}$ with no information about the $T^{tt}$ or $T^{tr}$ components.  However we see that $T^{tt}=\rho\gamma^{2}=-1/\epsilon C$ so that this corresponds to a net \textit{negative} mass density.  The vanishing to all orders of the $T^{tr}$ terms implies that there is an equal mixture of particles moving inwards and outwards.  This seems like a dishearteningly unphysical solution.  We have not, however, exploited the fact that $\epsilon$ may be small but also have some small spatial variation at a given time.  

Our goal is going to be to piece together a set of infinitesimal mass shells with possibly large boost parameters and even allow for finite time-frozen vacuum annuli between them.  This is a necessary freedom because we could form a black hole and later allow shells of mass that are large enough they time-freeze at a finite distance from the inner black hole giving one of a larger size.  Consider the modified metric 
\begin{equation}
g_{\mu\nu}^{(\text{mod})}=
\begin{pmatrix}
\epsilon r+a & 0 & 0 & 0 \\ 0 & -F(r) & 0 & 0\\0 & 0 & -C & 0\\0 & 0 & 0 & -C\sin^{2}(\theta)
\end{pmatrix}
\end{equation}
This gives the modified Einstein tensor
\begin{equation}
G_{\mu\nu}^{(\text{mod})}=
\begin{pmatrix}
-\frac{\epsilon r+a}{C} & 0 & 0 & 0 \\ 0 & \frac{F(r)}{C} & 0 & 0\\0 & 0 & \sim\epsilon & 0\\0 & 0 & 0 & \sim\epsilon
\end{pmatrix}
\end{equation}
By choosing $a<-\epsilon r_{1}$ for a solution in the neighborhood of a radius $r_{1}$ we have a positive mass density solution.  We see that small changes in the already tiny $g_{tt}$ value makes a large change in the energy density but negligible changes in the $T_{ij}$ terms (where roman indices correspond to spatial coordinates).  Ultimately, we have little interest in the particulars of the $g_{tt}(r)$ value on the interior other than it be very small.  Having the freedom to ensure it stays small and can give any $\rho\gamma^{2}$ value desired is sufficient for our purposes.  

We also require a class of vertical-cone time-frozen vacuum solutions which are easily found to be 
\begin{equation}
g_{\mu\nu}^{(\text{vac})}=\alpha
\begin{pmatrix}
\epsilon & 0 & 0 & 0 \\ 0 & -1 & 0 & 0\\0 & 0 & -r^{2} & 0\\0 & 0 & 0 & -r^{2}\sin^{2}(\theta)
\end{pmatrix}
\end{equation}
where $\alpha$ is a scalar parameter.  

For the outer part of the metric we choose the standard isotropic coordinate solution as the vertical-cone vacuum solution we will match to:
\begin{equation}
g_{\mu\nu}^{(\text{outer})}=
\begin{pmatrix}
\left(\frac{1-m/2r}{1+m/2r}\right)^{2}  & 0 & 0 & 0 \\ 0 & -(1+m/2r)^{4} & 0 & 0\\0 & 0 & -r^{2}(1+m/2r)^{4} & 0\\0 & 0 & 0 &  -r^{2}\sin^{2}(\theta)(1+m/2r)^{4}
\end{pmatrix}
\end{equation}
where $m$ is the mass parameter that gives the observed asymptotic Newtonian field.  At the Schwarzchild radius $r_{0}=m/2$ this takes the form
\begin{equation}
g_{\mu\nu}^{(\text{outer})}|_{r_{0}}=
\begin{pmatrix}
0  & 0 & 0 & 0 \\ 0 & -2^{4} & 0 & 0\\0 & 0 & -2^{2}m^{2} & 0\\0 & 0 & 0 &  -2^{2}m^{2}\sin^{2}(\theta)
\end{pmatrix}
\end{equation}

We now need to piece these together into a persistent vertical-cone global solution.  It is sufficient to match the metrics using the $(r,\theta)$ subsectors
\begin{equation}
\tilde{g}_{ij}=
\begin{pmatrix}
g_{rr} & 0\\0 & g_{\theta\theta} 
\end{pmatrix}
\end{equation}

As a first example, let us consider a black hole of radius $r_{0}$ and mass parameter $m=2r_{0}$ with a spherical interior vacuum region of radius $r_{1}=\beta r_{0}\lesssim r_{0}$.  Matching metrics $g^{(outer)}, g^{(mod)}, g^{(vac)}$ we have 
\begin{align}
C&=4m^{2}\\
\alpha&=\frac{2^{4}}{\beta^{2}}
\end{align}

\begin{align}
g_{rr}&=\begin{cases}
   -\alpha      & \text{if } r < r_{1} \\
   -F(r)      & \text{if } r_{1}<r < r_{0} \\
   -(1+m/2r)^{4}         & \text{if }  r_{0}<r
  \end{cases}\\
g_{\theta\theta}&=\begin{cases}
   -\alpha r^{2}     & \text{if } r < r_{1} \\
   -C      & \text{if } r_{1}<r < r_{0} \\
   -r^{2}(1+m/2r)^{4}         & \text{if }  r_{0}<r
  \end{cases}\\
\end{align}
where 
\begin{align}
\frac{F(r)}{2^{4}}&=\left(1-\frac{1+\beta^{-2}}{1-\beta}\right)+\frac{1+\beta^{-2}}{1-\beta}\frac{r}{r_{0}}\\
&=\frac{1+\beta+\beta^{2}}{\beta^{2}}-\frac{1+\beta}{\beta^{2}}\frac{r}{r_{0}}
\end{align}
The discontinuities in the second derivatives will produce undesirable singular source contributions but in the limit of small $dm$ and a quasicontinuum of many such small layers we can diminish these and build up many different time-frozen histories with various density distributions corresponding to the same static external field of the body.

From the above metric we see that our knowledge about the distribution of mass energy is summed up in $T_{rr}=8\pi F(r)/C$ and there are many functions $F(r)$ that give the same outer gravitational field.  Let us assume we start will a dilute ball of dust with large radius such that the asymptotic field is that of a classical point mass $m$.  In this case the mass of the cloud must also be $m$.  As the dust falls inwards, assuming it is sufficiently uniform so that the radiation losses are negligible, it creates a time-frozen ball and the external fields relax to the above case with no change in the field at infinity.  Assuming there are $N$ particles with fixed mass then the mass density is always well defined yet it has no important contribution to the final interior solution.  Only the a function of it and the final kinetic information, through $\rho \gamma^{2}$, is relevant.  The initial ball could have been a single uniform region or a sequence of disjoint shells that arrived at different times.  This is reflected in the final function $F(r)$ but many other mass distributions with other velocity fields give the same one.  The one way we can extract this information again is to somehow undo the black hole by strong external fields that strip away layers in the sequence they fell in.  In this case we can pull out the various particle types and densities in reverse order.  

It is often stated that, while the surface are of a black hole is well defined, the radius and volume are not.  The various singular solutions give such conclusions.  We are now in a position to label all such solutions ``transfinite'' in that they presume no interference in their formation for all of the external time.  For time changing nonlocal configurations, one can always dispute the meaningfulness of geometric quantities like a volume but, once it is static with respect to an asymptotically fixed spacetime, the choice of spacelike foliations are no longer relevant and the volume and other geometric quantities are well defined.  
From our time-frozen solution, we see that the radius is $l=\int_{0}^{r_{0}} dr\sqrt{-g_{rr}(r)}$.  In the above example we have the result
\begin{align}
l&=l_{1}+l_{2}\\
&=\int_{0}^{r_{1}}dr\sqrt{\alpha}+\int_{r_{1}}^{r_{0}}dr \sqrt{F(r)}\\
&=4 r_{0}+\frac{8}{3}\frac{1-\beta^{3}}{\beta(1+\beta)}r_{0}\\
&=(2m)\left(1 +\frac{2}{3}\frac{1-\beta^{3}}{\beta(1+\beta)}\right)
\end{align}

The spatial measure is  
\begin{align}
\sqrt{|g_{ij}|}=\sqrt{-g_{rr} g_{\theta\theta}g_{\phi\phi}}=\begin{cases}
   \alpha^{3/2}r^{2}\sin(\theta)      & \text{if } r < r_{1} \\
   \sqrt{F(r)}C\sin(\theta)      & \text{if } r_{1}<r < r_{0}
  \end{cases}\\
\end{align}
so the volume is given by 
\begin{align}
V&=\int_{0}^{r_{0}}\int_{0}^{2\pi}\int_{0}^{\pi} dr d\theta d\phi \sqrt{|g_{ij}|}\\
&=4\pi \left(  \frac{1}{3}\alpha^{3/2}r_{1}^{3}     + C l_{2}    \right)\\
&=4\pi \left(  \frac{2^{3}}{3}m^{3}     + 4 m^{2} l_{2}    \right)\\
&=4\pi m^{3} \left(  \frac{2^{3}}{3}    + \frac{2^{4}}{3}\frac{1-\beta^{3}}{\beta(1+\beta)}   \right)\\
&=\frac{4}{3}\pi (2m)^{3} \left(  1    + 2\frac{1-\beta^{3}}{\beta(1+\beta)}   \right)
\end{align}
It is interesting that this radius is not uniquely fixed by the mass of the body (defined by asymptotic fields) but depends on the history of its formation.  We already saw that the particle number that formed the body is not determined by the external field.  Similarly, the volume cannot be determined by outside observers who did not watch it form unless they have a way to systematically rend the system apart.  This assumes there is no angular momentum or charges in any of the layers that went into its creation.  As we shall see below, these may give telltale signs of inner structure.

\section{Persistent Boundary Effects}
The location of the degenerate boundary of a system, where $g_{tt}\rightarrow0$, is where the time evolution rate of physical fields vanishes in our infalling solution.  It creates a kind of boundary condition on the exterior metric.  The inner core defined by this boundary contains all the history of angular momentum, charge, etc.\ that have been trapped during the infall and are now practically out of reach of the external world.  Let us now consider some aspects of this boundary and its implications. 

When a black hole moves into regions of slowly changing external field we expect it to move and respond in a semi-Newtonian fashion.  From the standpoint of our infalling masses, this raises an interesting paradox.  They must have finite coordinate translational velocity yet have net velocity approximately $c(x)$ as defined by the local (now tilted) cone so be an interesting mix of inwards infinitesimal and finite translational motion.  If we now view the exterior field as due to a single source, and consider the ``boosted'' situation as viewed from far away, then  the black hole is at rest and the external fields of the moving object must penetrate it in finite external observer time.  It things are truly time-frozen on the interior, how can changes in external fields ever manifest on the interior?  The resolution is in noting that our notion of ``time-frozen'' is a bit naive.  

We have focused on solutions with zero light cone tilt so that as they narrowed the maximum coordinate velocity had to vanish.  As we mentioned, nonzero currents will require us to consider tilted and deformed light cone solutions.  A translating black hole will be a finite volume region of degenerate forward tilting cones.  As long as these cones don't ``overtilt'' and destroy our spacelike foliation things are fine.  By our general causality condition, the evolution is always chosen so that this is so.  This can always be done since we have fixed an asymptotic flat space to act as our preferred viewers and as their time $t_{asyp}\rightarrow\infty$ their observations of the particles interior to them determine the rate that their time should be allowed to advance.  In the case of accelerating black holes where holes are moving into new external field regions, the cone structure must become nondegenerate by widening out so that information can penetrate the bulk of the body.  How to realize this is not obvious and the latter part of this paper is dedicated to a proposed method.

Consider the case of a rotating black hole.  It is still debated how to properly patch the Kerr metric  outside the even horizon to a physical interior solution but it does provide a standard to investigate rotating black holes.  Now that we suspect that black holes may have richer properties than the no-hair conjecture claims, we may wonder how varied the boundary regions and external fields can be.  Let us consider a simple perturbative example.  Consider the previous case of infalling dust but now include a ring of of particles mass $dm$ that has a net angular velocity $\omega R$ and angular momentum $dL$.  This is well defined since the cloud is dilute and far away and it generates a gravitomagnetic field at infinity.  As it falls in to the surface, it gains angular velocity to preserve the angular momentum until relativistic corrections and time-freezing reduces it to zero.  Since the system remains axially symmetric, angular momentum is conserved, in particular, the fields at infinity are unchanged between the initial state and the final relaxed state.  The observers at infinity either see the magnetic field persist or die out.  If the ring of mass is charged a similar argument holds for the induced magnetic field.  This leads us to believe that magnetic fields are indeed possible for black holes.  Here is a situation where the gravitational field is dominated by a nonrotating part and the rotation is a mere perturbation.  General rotational solutions allow currents that are not infinitesimal.  In such a case, the magnetic flux from such a circulation has even less reason to vanish.

Consider a second example beginning with our time-frozen dust solution.  Let two charges $+q$ and $-q$ fall inwards to opposite poles of the ball.  These generate magnetic fields which vanish in the limit of adiabatic descent of the charges to the boundary.  Since these are discrete entities they cannot diffuse out to form a uniform charge distribution over the surface.  The final solution is a charge distribution dominated by the dipole field at infinity corresponding to a dipole moment $d\sim 4ql$.  The presence of strong persistent multipole fields would certainly have a strong effect on the external dust and gas about a black holes so is ultimately an experimental observable however, the numerical simulation methods available today all assume centrally singular solutions which preclude such dynamics.  After discussing a reformulation of GR below, we will return to this point.

\section{The Hidden Flat Background Formulation}  
By our previous considerations we see that, for an initially nearly flat space coverable with one chart, this coverability property should persist for all times even when the space develops large curvature.  Moving black holes can have light cones that ``overtilt'' and become degenerate but maintain a well defined sense of the future through the half-space given by $t>t_{0}$ for a set of slices given by $\mathbb{R}^{3}\times \mathbb{R}^{1}$.  
In the 3+1 Hamiltonian decomposition \cite{ADM} approach this implies we should only choose the lapse and shift functions so that this remains valid.  This removes the usual concerns about introducing more charts with time due to evolution or rotation and having to consider only wisely chosen evolving charts.  

The price we pay is that black holes become finite volume regions of near degenerate metric.  Numerically, this situation rapidly becomes unstable.  A similar situation exists for shock waves where singular behavior ruins the use of numerical approaches to the pdes of hydrodynamics.  In this case, conservation laws are used to derive the Rankine-Hugoniot conditions to help evolve the shock while standard numerical methods handle the rest of the flow.  If such an analogous approach is to work in GR we need some local conservation laws to manage the evolution at the degenerate boundary.  It is not clear such an approach will be fruitful.  GR is notoriously stingy with conservation laws.  The reason for this is that the Killing vectors corresponding to the usual ten symmetries of classical physics are generally missing due to the broken symmetries in the metric from nontrivial configurations.

It would be highly desirable if we could rewrite the equations in terms of fields where no such functions need to be chosen and the gravitational field evolved according to hyperbolic pdes on a flat background.  Ultimately, we only need local conservation laws.  These already exist in GR with the usual metric approach and we will discuss how to use them in this context.  
One can object to the ``physicality'' of such an approach but, if the spacetime can be so foliated for all values of the time of our asymptotic observers, then we have tools to begin such a program.  The ``physical'' meaning of it can be discussed and reconsidered later.  For now, we leave the general existence of the foliation as given and proceed.  

To make this transition 
we define the flat space Lorentz metric $\eta^{\mu\nu}$ and the gravity field $h^{\alpha\beta}$ implicitly by the association $g_{\mu\nu}=\sum_{\alpha} h_{\mu\alpha} \eta_{\alpha\nu}=h\circ\eta$ where $g$ is the GR metric.\footnote{This is not the usual small perturbation of the flat background metric indicated by $g=\eta+h$.  Here the flat background is considered the governing geometric driver of causality and the field $h$ is a classical tensor field of positive determinant that introduces nonlinear effects.}  For vertical-cone solutions, this definition ensures that the field $h$ has positive signature $[+,+,+,+]$.  More generally, the field $h_{\mu\nu}$ has a positive determinant.  
We now define the field equations as tensor equations in the flat space metric $\eta$.  The ``geometric'' coordinate transformations of the tensor fields are now downgraded to a more mundane class of coordinate independent gauge transformations on the gravity field $h$.  The notion of such a background approach is generally rejected because it does not allow changes in the number of charts as the spacetime evolves.  By the previous sections, this is not necessary.  The role of $\eta$, although a flat space metric, will be upgraded to give the new ``true'' underlying geometry of the space for variational purposes.  

Here we will give a quick survey of the approach then make it explicit in what follows.  Let us rewrite the equations of GR explicitly in terms of the metric, its inverse and other fields.  We start with the action
\begin{align}\label{action}
S=\int d^{4}x \mathcal{L}\sqrt{|g_{..}|}=\int d^{4}x \left(\mathcal{L}_{\text{fields}}(\varphi, A_{\mu},g_{\mu\nu}\ldots)+\frac{1}{16\pi}R(g_{\mu\nu})\right)\sqrt{|g_{..}|}
\end{align}
where $R$ is the Ricci scalar defined in terms of the metric $g_{\mu\nu}$.  (The very act of using an action to derive the field equations in GR has conceptual problems \cite{York1}.  Boundary conditions at infinity are generally not ignorable but using a flat background gives us a solid underlying structure that eliminates these as concerns for setting up a well defined variation.)  The Lagrangian is a function of $g_{\mu\nu}$, its inverse $g^{\mu\nu}\equiv g^{\#}_{..}$, and derivatives of them.  Covariant derivatives are with respect to the metric $g_{\mu\nu}$ which we write as $\nabla(g_{\mu\nu})$.  Now we make the association $g_{\mu\nu}\rightarrow (h\circ\eta)_{\mu\nu}$ where $\eta_{\mu\nu}$ is, for now, a constant metric of an underlying flat spacetime and $h_{\mu\nu}$ is a two component tensor field.  
We next reconsider our raising and lowering rules of indices defined by using the metric $g_{\mu\nu}$.  Our goal will be to convert all fields to lower index objects (where the notion of ``$\eta$-index'' and ``$g$-index'' objects coincide) and have the only ``upper index'' object in $\mathcal{L}$ be $\eta^{\mu\nu}\equiv\eta^{-1}$ itself.  

Once this is done we redefine the raising and lowering rules be done with the flat space metric $\eta_{\mu\nu}$ and its inverse $(\eta^{\#})^{\mu\nu}$ even though, in general coordinates, these may be very nontrivial location dependent objects.  In the general transformation of coordinates, we have to define the metric to define the tensor transformation rules for differentials and the measure for invariant integrals.  The measure-like factor in $\mathcal{L}$ can be decomposed as such: $\sqrt{|g|}\rightarrow\sqrt{-\det(\eta)}\sqrt{\det(h)}$ where $\sqrt{-\det(\eta)}$ is the integration measure and the $\sqrt{\det(h)}$ factor gives a dynamic nonquadratic contribution to the action.  For the constant metric $\eta_{\mu\nu}=\text{diag}(1,-1,-1,-1)$ the derivatives in expressions such as $h_{\mu\nu,\alpha}$ are interpreted as coordinate derivatives.  When we adopt curvilinear coordinates on the flat background space these become covariant derivative operators in the new (but equivalent) metric functions $\eta'$ as $\partial_{\alpha}\rightarrow\nabla_{\alpha}(\eta')$.  
Instead of deriving the equations of motion by variation of the action by the metric $\delta S/\delta g^{\mu\nu}$ we use $\delta S/\delta h^{\mu\nu}$.  Since the value of the action is unchanged and $\eta_{\mu\nu}$ describes a constant uniform background the resulting dynamics must be the same.  

The usual action approach by variation of $g$ yeilds the Einstein equations
\begin{align}\label{Einstein}
G_{\mu\nu}=R_{\mu\nu}-\frac{1}{2}g_{\mu\nu}R=8\pi T_{\mu\nu}
\end{align}
where the Ricci tensor and Riemann scalar are defined in terms of the connection 
\begin{align}
R_{\alpha\beta}=\partial_{\rho}\Gamma^{\rho}_{\beta\alpha}-\partial_{\beta}\Gamma^{\rho}_{\rho\alpha}+\Gamma^{\rho}_{\rho\lambda}\Gamma^{\lambda}_{\beta\alpha}-\Gamma^{\rho}_{\beta\lambda}\Gamma^{\lambda}_{\rho\alpha}
\end{align}
and the connection is defined in terms of the metric
\begin{align}
\Gamma^{\alpha}_{\beta\gamma}=\frac{1}{2}g^{\alpha\mu}\left( \partial_{\gamma}g_{\mu\beta}+\partial_{\beta}g_{\mu\gamma}-\partial_{\mu}g_{\beta\gamma} \right)
\end{align}
The stress tensor source term here is given by 
\begin{align}
T_{{\mu\nu}}^{(g)}=\frac{-2}{\sqrt{-\det(g)}}
\frac{\delta (\sqrt{-\det(g)}\mathcal{L}_{\text{fields}}) }{\delta g^{{\mu\nu}}}
\end{align} 
so it is a function of the nongravitational fields $A^{\mu}\ldots$ and $g_{\mu\nu}$.  Here we define $\delta/\delta g^{\mu\nu}$ in terms of $g_{\mu\nu}$ though the constraint $g_{\alpha\mu}g^{\mu\beta}=\delta_{\alpha}^{\beta}$.

Now we make the following insertion $g=\eta\circ h$ or, equivalently, $g_{\mu\nu}\rightarrow \sum_{\alpha}\eta_{\mu\alpha}h_{\alpha\nu}$ and define the auxiliary field $\tilde{h}_{\mu\nu}\equiv \sum_{\alpha}\sum_{\beta}(h^{-1})_{\alpha\beta}\eta_{\alpha\mu}\eta_{\beta\nu}$ so that $g^{\#}=(\eta^{-1}\circ\tilde{h}\circ\eta^{-1})\circ\eta^{-1}$.  Einstein's equations eqn.\ \ref{Einstein} are now in a form where the only raised indices are from the background flat $\eta$ metric and all the derivatives are ordinary derivatives.  We now let the true ``geometric'' metric be interpreted as $\eta^{\mu\nu}\rightarrow\eta^{\mu\nu}(X)$ which now corresponds to the same flat spacetime geometry with arbitrary coordinates and replace the ordinary derivatives with $\eta$-covariant ones.  For example, an action on an ``$\eta$-vector'' $A_{\mu}$ now takes the form
\begin{align}
\partial_{\alpha}A_{\mu}\rightarrow\nabla_{\alpha}(\eta)A_{\mu}=\partial_{\alpha}A_{\mu}-\Gamma(\eta)^{\nu}_{\alpha\mu}A_{\nu}
\end{align}
where the $\eta$-connection is defined by 
\begin{align}
\Gamma(\eta)^{\alpha}_{\beta\gamma}=\frac{1}{2}\eta^{\alpha\mu}\left( \partial_{\gamma}
\eta_{\mu\beta}+\partial_{\beta}\eta_{\mu\gamma}-\partial_{\mu}\eta_{\beta\gamma} \right)
\end{align}
The auxiliary field $\tilde{h}_{\mu\nu}$ must be chosen to satisfy $\eta^{\mu\alpha}\eta^{\nu\beta}\tilde{h}_{\alpha\beta}h_{\nu\sigma}=\delta^{\mu}_{\sigma}$ for all spacetime.  This immediately gives an implicit (first order) evolution equation for $\tilde{h}$ from $\partial_{t}\{(\eta^{-1}\circ\tilde{h}\circ\eta^{-1})\circ h\}=0$.  The initial data for $h$ includes $\dot{h}$ but since the eom of $\tilde{h}$ is fixed so is that of its first time derivative.   This gives a set of evolution equation that agree with Einstein's on a flat background in a ``nongeometric'' form that \textit{is} manifestly invariant with respect to arbitrary coordinate changes on the flat background.  The gauge invariance we associate with GR is now in the subset of transformations of the nonmetric function $g_{\mu\nu}$ that is consistent with the global slice structure implied by the background and correspond to coordinate changes in the usual geometrodynamic formalism.  The coordinate invariance we expect of relativity still exists but does not treat the fundamental fields as tensorial objects.  However, as observed by local observers built of fields described by such a Lagrangian, nothing has changed.  He can still rearrange rods and clocks freely to give new physical approximations to coordinates that he directly perceives.  These are not simple transformations of coordinates in the ``$\eta$-picture'' where the transformations now involve nongeometric gauge transformations of the fields.  The distinction is immaterial and this gives an example of the ``field-relativity'' discussed earlier.

On variation by $g^{\mu\nu}$ of the action in eqn. \ref{action} we get $G_{\mu\nu}(g)$ which, by Bianchi identities gives $g^{\alpha\mu}\nabla_{\alpha}(g)G_{\mu\nu}(g)=0$ and a term $T^{\mu\nu}$.  This implies the variation of the $\mathcal{L}_{\text{fields}}$ part of the equation gives a conserved stress tensor.  We can rewrite this in terms of connections as 
\begin{align}
\partial_{\mu}T^{\mu\nu}+\Gamma_{\alpha\mu}^{\mu} T^{\alpha\nu}+\Gamma_{\alpha\mu}^{\nu} T^{\mu\alpha}=0
\end{align}
where $T$ and $\Gamma$ are written in terms of the background $\eta$, the gravity field $h$ and the other fields in $\mathcal{L}_{\text{fields}}$.  
Without Killing vectors to convert this into a conserved global expression we still have a local result that gives us the value of $\partial_{t}T^{t\nu}$, in other words, the energy and momentum fluxes with the ``source'' terms of the form $\Gamma T$.  Since the limiting cone degeneracy does not generally create divergences in the $\Gamma^{a}_{bc}$'s, this gives a local criterion to attempt to evolve the degenerate cones.

Having a local set of conservation laws is a promising first step towards a treatment of evolution at the degenerate boundary.  In the time-frozen region we have to keep track of the stored stress-energy and the details of fields and particle types stored there.  This means we will need to keep both $T^{\mu\nu}$ and $T_{\mu\nu}$ information (since one cannot be derived from the other in this region) and particular details of the fields.  At such a degenerate boundary, all the physical conservation laws: energy, momentum, charge, lepton number etc.\ will have to be respected in the local evolution equations for the metric and fields.  The narrow cone structure opening up in the interior implies the presence of a lag in the response time of a gravitional signal crossing the black hole suggests that acceleration may be accompanied by large deformations to its shape; an interesting question for future work.  

From our observations that charge distributions can get frozen in nontrivial patterns to the surface of the horizon, one has to wonder what this means for the standard numerical approaches to GR.  Currently we use versions of the ``punctured disk'' method where a singular interior is fitted to the external field.  Based on previous arguments it seems that we may have been pushing data down into the hole too fast and over simplifying the external field.  While the time-frozen picture is not unknown to many researchers, the prevailing attitude is that ``for all intents and purposes'' both approaches must be equivalent.  It would certainly be interesting to develop this alternate picture further for a true comparison.  

%As a final example of the benefits of this approach consider the question of ``positive energy'' in GR.  This is a famously difficult problem \cite{Schoen}.  For our approach this becomes somewhat trivial for diffuse low curvature initial data.  We can consider all the initial particles at a time before spacetime had any dense clusters.  The mass of a particle is a combination of the mass parameters in their governing field equations and the local gravitational field each particle makes.  If the mass parameters are all positive then it becomes an easy exercise here to determine the positive energy of the universe.  Since this is strictly conserved in our $\eta$-space coordinates, it is positive for all times in any spacelike slice foliation.  In our time frozen solution we chose all the points inside the outer boundary to be time frozen as well.  We could have patched the coordinates to an interior where this was not the case but, trials indicate that this leads to negative energy contributions.  As a hypothesis, we suggest that, in universe with all positive mass particles, the interior of any black hole is time frozen relative to external observers.  

\section{Conclusions}
We have elucidated the time frozen picture of a black hole with specific solutions and demonstrated its necessity and inequivalence with the Lagrangian observer picture.  The interior and persistent external fields have been shown be much more exotic than the no-hair conjecture suggests and this has led to a probable conflict with the popular punctured-disk approach to numerical simulations.  Having a second approach to analyze the external fields of black holes under evolution would be either a comforting situation in confirming them or a revolutionary one in refutation.  It is unclear how novel this perspective really is since the idea of the ``frozen star'' is very old and may date back to the first infalling particle solutions for black holes.  Nevertheless, it has implications that are destructive to many popular and persistent ideas in GR and should be a foil for future work.  

Black holes periodically get attention in the popular media.  Unfortunately, this often takes the form of exotic statements about Hawking radiation and quantum fluctuations near or inside the event horizon.  In a dynamical sense, event horizons are among the most boring places in the universe.  Aside from the external nudges they obtain from external sources that alter the black hole's motion, the horizon demarks a rather dull repository of past history held in suspension until a collision or near grazing event with another such body occurs.  The degeneracy in the metric at such locations makes the way forwards murky but having rigid local conservation laws, even at the expense of less appealing equations and more physically opaque variables, suggests a resolution.  

Some have considered the distinction between the singular black holes and frozen stars as pedagogical.  It is clear from this discussion that this is not the case.  The external fields of the general frozen star is significantly different and the consequences for black hole collisions may not be equivalent to the numerical simulations derived from the ``punctured disk'' method, certainly for black holes that are not primordial.  Furthermore, the rather evolved body of work on trapped surfaces and the inevitability of black hole formation in some circumstances is now potentially irrelevant.  It brings up further questions on the reality of Hawking radiation and the no-hair conjecture.  The problems of entropy and information vanishing into black holes and the relevance of cloaking of singularities may also be undefined as problems for the subject.  

Aside from its necessity for consistency, an obvious benefit of such a global point of view is in quantum field theory.  The structure of the formalism seems to strongly depend on a fixed background.  There has been work to get around this but how to carry field theory results across event horizons or put all the forces of nature on a uniform foundation remains unclear.  Removing the geometric role of gravity introduces some inelegance in the formulation of the subject (if not as much in actual practice) but allows a new approach to these problems and, furthermore, allows us to then ask why gravity seems to have a special geometric role in its interactions with other fields.  The singular nature of propagators on the event horizons may provide a mechanism to enforce a consistent cone structure for all fields and drive nature to select one particular field (or combination of them) to act in a emergent geometric fashion.  A variational principle in which \textit{all} fields and derivatives of them are coupled together with varying coupling functions might give general motion towards such a state without any a priori favoritism or asymmetry in the fields.

\appendix
\section{Stationary Criteria}
We can always take a given metric function and derive a corresponding stress tensor that satisfies Einstein's equations.  The corresponding stress may not be physical e.g.\ $\rho$ defined by the density in the time-like direction may be negative, its corresponding currents may not correspond to a c-limited velocity or the equation of state violate some thermodynamic constraints.  In turn, the equations of motion depend on the metric so any stationary solution must give constant currents.  The wrinkle in this is that the coordinates of the space can be completely general and time dependent.  Even though GR is generally framed in terms of differential geometry we now argue that 
the abstract notion of point in manifold geometry seems insufficient to characterize the notion of ``location'' generally enough to define what we mean by a stationary solution.  What defines a point in spacetime?  We have coordinate labels but we can impose flows of the underlying points which have no physical meaning.  This could transform a stationary solution for one set of coordinates and general point labels to a periodic or seemingly chaotic one.  

To define a stationary state generally, we do not expect that all the local observers see the same local reality for all times so no local observer criterion exists. A stationary state is one where all the other fields (or more generally, their complete gauge independent features) and the metric are locally invariant for \textit{some} specification of points on the spatial manifold and some evolution of them so that these hold.  
This allows observers to undergo time changing effects where the overall solution undergoes no change at all.  Of course, we opt for a set of coordinates and underlying point structure that is fixed for our examples and a gauge choice that has no time dependence.  This then reduces the condition for a stationary state to the statement that the fields and metrics are constant at each point.  The temporal coordinates can then be chosen to be separable from the spatial ones.  The rate of evolution of observed time will vary with location but none of the usual problems of coordinate chart redefinition in numerical relativity arises.  

A manifold is defined as a set of abstract points with a local $\mathbb{R}^{n}$ topology.  We then build an atlas of charts to cover these and provide overlap conditions.  Our problem is that the points themselves should have no meaning much less a permanent reality.  Alternately, consider a set of bounded subsets of $\mathbb{R}^{n}$ charts and assign some overlap conditions.  These charts can have any relation to each other.  Since time is part of this collection the size and motion of the spatial restriction of the charts can change arbitrarily (as long as it is smooth).  These coordinate values themselves can be the ``points'' and the associations for the overlaps give equivalence classes generated by them.  A physical definition of ``stationary'' could then be that the metrical relations between any ``infinitesimal'' physical rays sent in from infinity with the same initial data are the same.  If there are sources present in the spacetime then we would require these relations also be fixed between them.  Without such benchmarks it is problematic to try to specify relations between test particles on the interior since we have no way to position them among the moving charts at different ``times'' so that they are in the same ``position.''  The formalism of manifold theory tends to obscure this problem but it seems that some of the flavor Mach's observations on rotation remains important.  For the mathematical purist who sees physics as simply a problem to be encoded in mathematics these discussions may seem uninteresting.  It is that author's opinion that physics gone far in the direction of mathematical shortcuts to conceptual understanding and that a probable price of this is the kind of computation driven confusions that have long dogged this subject.  

%COGNATIVE DISSONANCE.

\end{document}